\documentclass[12pt]{article}
\usepackage{amsmath}
\usepackage{amsfonts}
\usepackage{amssymb}
\usepackage{graphicx}

\input{tcilatex}
\begin{document}

\begin{titlepage}
\ \\
\begin{center}
\LARGE
{\bf
A Protocol for\\
 Quantum  Energy Distribution
}
\end{center}
\ \\
\begin{center}
\large{
Masahiro Hotta
}\\
\ \\
\ \\
{\it
Department of Physics, Faculty of Science, Tohoku University,\\
Sendai, 980-8578, Japan\\
hotta@tuhep.phys.tohoku.ac.jp
}
\end{center}
\begin{abstract}
In this paper,  a protocol called quantum energy distribution(QED) is  proposed 
in which  multi-parties can simultaneously extract positive energy on average 
 from spin chains by use of common secret keys shared  by an energy supplier. 
QED is robust against impersonation. 
An adversary, who does not have common secret keys and attempts to get energy, 
cannot obtain but give energy to spin chains. 
Total amount of energy transfer gives a lower bound of residual energy of a local cooling process by the energy supplier.   
\end{abstract}
\end{titlepage}

\bigskip

\section{Introduction}

\ \newline

Quantum teleportation \cite{qt} transfers any unknown quantum state to
distant places only by local operations and classical communication (LOCC).
It has attracted much attention and been widely investigated. Today, it is
considered \cite{qtr} as a crucial building block of quantum communication.
Recently, a new protocol named quantum energy teleportation (QET) in
spin-chain systems has been proposed\cite{hotta1}, which transports energy
from one location to another only by LOCC. Entanglement of \ spin-chain
ground states plays an essential role to realize QET.

The protocol for QET \cite{hotta1} has been proposed for general spin chains
with entangled ground states. Even before the advent of QET, spin chain
systems have been hot topics of quantum information theory, because it is
possible to apply it to short transmission of quantum information\cite{stqi}%
. It is also known \cite{rev}that spin-chain entanglement is important to
shed new light on complicated physical properties of ground states.

In the QET protocol, a receiver of classical information from an energy
supplier extracts positive energy from the ground state, accompanied by
generation negative energy density in spin chain systems. Here, the zero of
energy in the system is naturally defined by a value of the ground state.
Though the concept of negative energy density is not so familiar to quantum
information theory \ and quantum communication, \ it has been investigated
in relativistic field theory for long time \cite{BD}. \ Detailed analysis
for the spin chains can be seen in \cite{hotta1}.

In this paper, an extended protocol is proposed, in which many authenticated
consumers are able to simultaneously extract energy from the ground state by
use of common secret keys shared by an energy supplier. Let us later call
the protocol quantum energy distribution (QED). QED shows robustness against
impersonation. Let us imagine that an illegal consumer appears, who does not
have common secret keys and attempts to get energy from spin chains. Then we
can conclude that the adversary does not obtain but give energy to the spin
chains. We also notice that total amount of energy transfer of QED is
related with local cooling. Local cooling is a short-time process in which
energy is extracted from an excited system only by local operations at a
certain site, without use of global time evolution generated by the system
dynamics. In general, local cooling is unable to extract all energy of the
excited system and residual energy remains in the system. The total amount
of energy distributed via QED gives a lower bound of that residual energy of
a supplier's local cooling for an excited state. We also analyze QET and QED
protocols in the Ising spin chain system. Amount of energy transmission is
evaluated depending on distance from the supplier.

We confine our attention to short-time-scale processes in which dynamical
evolution induced by the Hamiltonian is negligible. Meanwhile let us assume
that classical communication between qubits can be repeated many times even
in the short time interval.

This paper is organized as follows. In section 2, we briefly review QET. In
section 3, extending QET, a QED protocol is proposed. In section 4, we
discuss a relation between QED and local cooling. In section 5, we analyze
the Ising spin chain system and demonstrate the QED protocol. In the final
section, conclusion is given.

\bigskip

\bigskip

\section{Brief Review of QET}

\ \newline

\bigskip

In this section, we shortly review QET. Detailed explanation is seen in \cite%
{hotta1}, including negative-energy physics of spin chains. Let us consider
a very long spin chain system with Hamiltonian given by

\begin{equation*}
H=\sum_{n}T_{n},
\end{equation*}%
where $T_{n}$ is the $n$th site energy density operator. In order to capture
the essence of QET, let us focus on the nearest neighborhood interaction
case. The operator $T_{n}$ is Hermitian and take the form of

\begin{equation*}
T_{n}=\sum_{\gamma }\prod_{m=n-1}^{n+1}O_{m}^{(n,\gamma )},
\end{equation*}%
where $O_{m}^{(n,\gamma )}$ is a local Hermitian operator at site $m$. The
ground state$|g\rangle $ is an eigenstate with the lowest eigenvalue of $H$.
When we do not take account of gravitational interaction, absolute values of
energy is irrelevant and just difference of values makes sense. Hence,
subtracting constants from energy density and the Hamiltonian, we obtain the
following relations without loss of generality.

\begin{equation}
\langle g|T_{n}|g\rangle =0,  \label{2}
\end{equation}%
\begin{equation}
H|g\rangle =0.  \label{1}
\end{equation}%
Due to Eq.(\ref{1}), the Hamiltonian becomes nonnegative:

\begin{equation*}
H\geq 0.
\end{equation*}

In many models, $|g\rangle $ is a complicated entangled state. Using the
entanglement, Alice who stays at site $n_{A}$ can transport energy to Bob at
site $n_{B}$ only by LOCC. Taking account of the nearest neighborhood
interactions, let us define localized energy operators of Alice and Bob as
follows.

\begin{equation*}
H_{A}=\sum_{n=n_{A}-1}^{n_{A}+1}T_{n},
\end{equation*}%
\begin{equation*}
H_{B}=\sum_{n=n_{B}-1}^{n_{B}+1}T_{n}.
\end{equation*}%
For later convenience, let us introduce several operators as follows. $U_{A}$
and $U_{B}$ are unitary Hermitian operators given by 
\begin{equation}
U_{A}=\vec{n}_{A}\cdot \vec{\sigma}_{n_{A}},  \label{ua}
\end{equation}%
\begin{equation*}
U_{B}=\vec{n}_{B}\cdot \vec{\sigma}_{n_{B}},
\end{equation*}%
where $\vec{\sigma}$ are Pauli vector matrices, $\vec{n}_{A}$ and $\vec{n}%
_{B}$ are three-dimensional real unit vectors. The operator $U_{A}$ can be
spectral decomposed into 
\begin{equation*}
U_{A}=\sum_{\mu =0,1}\left( -1\right) ^{\mu }P_{A}\left( \mu \right) ,
\end{equation*}%
where $P_{A}\left( \mu \right) $ is a projective operator onto the
eigenspace with an eigenvalue $\left( -1\right) ^{\mu }$ of $U_{A}$. $\dot{U}%
_{B}$ is time-derivative operator of $U_{B}$ defined by

\begin{equation*}
\dot{U}_{B}=i\left[ H_{B},~U_{B}\right] =i\left[ H,~U_{B}\right] .
\end{equation*}%
Next let us introduce two real coefficients as follows.%
\begin{equation}
\xi =\langle g|U_{B}^{\dag }HU_{B}|g\rangle >0,  \label{15}
\end{equation}%
\begin{equation}
\eta =\langle g|U_{A}\dot{U}_{B}|g\rangle .  \label{16}
\end{equation}%
Also define an angle parameter $\theta $ which satisfies

\begin{align*}
\cos \left( 2\theta \right) & =\frac{\xi }{\sqrt{\xi ^{2}+\eta ^{2}}}, \\
\sin (2\theta )& =-\frac{\eta }{\sqrt{\xi ^{2}+\eta ^{2}}}.
\end{align*}%
Finally we define a unitary matrix $V_{B}\left( \mu \right) $ \ for $\mu
=0,1~$as follows.

\begin{equation}
V_{B}\left( \mu \right) =I\cos \theta +i\left( -1\right) ^{\mu }U_{B}\sin
\theta .  \label{vb}
\end{equation}%
The parameter $\eta $ is important for QET. If $|g\rangle $ is separable, we
can generally prove that $\eta $ vanishes. As seen below, QET transports no
energy when $\eta =0$. Thus, in later discussion, we assume that $|g\rangle $
is an entangled state such that $\eta \neq 0$.

In order to perform QET, let us assume that Alice is a good distance from
Bob such that%
\begin{equation*}
\left\vert n_{A}-n_{B}\right\vert \geq 5,
\end{equation*}%
and that Alice and Bob share many copies of spin chain systems in the ground
state $|g\rangle $. Now let me explain the protocol explicitly. The protocol
is composed of three steps as follows.

\bigskip

(1) Alice performs a local projective measurement of the observable $U_{A%
\text{ }}$ for the ground state $|g\rangle $. Assume that she obtains the
measurement result $\mu $. She must input energy $E_{A}$ on average to the
spin chain system in order to achieve that local measurement.

\bigskip

(2) Alice announces to Bob the result $\mu $ by a classical channel.\bigskip

\bigskip

(3)Bob performs a local unitary operation $V_{B}\left( \mu \right) $ to his
qubit at site $n_{B}$, depending on the value of $\mu $.~Bob obtains energy
output $E_{B}~$on average from the spin chain system in this process.

\bigskip

\bigskip It is noticed \cite{hotta1} that the average input energy $E_{A}$
is evaluated as

\begin{equation}
E_{A}=\sum_{\mu =0,1}\langle g|P_{A}\left( \mu \right) HP_{A}\left( \mu
\right) |g\rangle >0.  \label{12}
\end{equation}%
A positive amount of energy $E_{B}$ is released to Bob's devices for the
operation $V_{B}\left( \mu \right) $ in step (3). $E_{B}$ is given by

\begin{equation}
E_{B}=\frac{1}{2}\left[ \sqrt{\xi ^{2}+\eta ^{2}}-\xi \right] .  \label{eb}
\end{equation}%
It is stressed that dissipation effect of transfered energy in channels can
be completely neglected for QET because we transmit only classical
information through a classical channel.

In the above analysis, it has been argued that Bob actually obtains energy
from the spin chain system. However, even after the last step of the
protocol, there exists energy $E_{A}$, that Alice first deposited to the
spin chain by herself. \ Then a natural question arises. Does Bob extract
positive energy without any cost ? This apprarent paradox can be resolved
from the viewpoint of entanglement breaking by Alice. Detailed explanation
is seen in \cite{hotta1}. It is concluded that, based on a pledge of $E_{A}$%
, Bob knowing classical information $\mu $ has borrowed $E_{B}$ in advance
from the spin chains. When global cooling induced by both long-time
evolution of the system and extraction of energy makes the state approaching
the ground state, \ the residual energy and negative energy $-E_{B}$ around
site $n_{B}$ are compensated.

In the QET protocol, classical channels for Alice to inform measurement
results are not assumed private and secure. Therefore, anybody can extract
energy from spin chains by listening to the measurement results announced by
Alice. In the next section, an extended protocol is proposed in which
legitimate multi-users can extract energy but illegal users are unable to
steal energy from spin chains at all.

\bigskip

\section{Quantum Energy Distribution}

\bigskip

\bigskip

In this section, a QED protocol is proposed, in which $M$ consumers $%
C_{m}~(m=1\sim M)$ can simultaneously extract energy from spin chains by use
of secret classical information sent by an energy supplier $S$. $\ $The
protocol is an extension of QET assisted by quantum key distribution(QKD).
Let us consider that $S$ stays at $n=0$. Assume that the spin chain is so
long that we are able to treat the number of sites as infinite and that the
entangled ground state has a very large (or divergent) correlation length.
Let us assume the sites of $S$ and $C_{m}$ are separated from each other
such that

\begin{equation*}
\left\vert n_{C_{m}}\right\vert \geq 5,
\end{equation*}%
\begin{equation*}
\left\vert n_{C_{m}}-n_{C_{m^{\prime }}}\right\vert \geq 5.
\end{equation*}%
Here let us introduce $U_{S}$ and $V_{C_{m}}$ as follows. 
\begin{equation*}
U_{S}=\vec{n}_{S}\cdot \vec{\sigma}_{0}=\sum_{\mu =0,1}\left( -1\right)
^{\mu }P_{S}\left( \mu \right) ,
\end{equation*}%
\begin{equation*}
V_{m}\left( \mu \right) =I\cos \theta +i\left( -1\right) ^{\mu }U_{m}\sin
\theta ,
\end{equation*}%
where $P_{S}\left( \mu \right) $ is a projective operator onto the
eigensubspace with an eigenvalue $\left( -1\right) ^{\mu }$ of $U_{S},$ 
\begin{equation*}
U_{m}=\vec{n}_{m}\cdot \vec{\sigma}_{n_{C_{m}}},
\end{equation*}%
and $\vec{n}_{S}$ and $\vec{n}_{m}$ are real normal vectors. The localized
energy operators for those consumers are given by%
\begin{equation*}
H_{C_{m}}=\sum_{n=n_{C_{m}}-1}^{n_{C_{m}}+1}T_{n}.
\end{equation*}%
Let us also define a time-derivative operator of $U_{m}$ as

\begin{equation*}
\dot{U}_{m}=i\left[ H,~U_{m}\right] =i\left[ H_{C_{m}},~U_{m}\right] .
\end{equation*}

Consider that supplier $S$ and any consumer $C_{m}$ share \ common secret
short keys $k$ for their identification, by which they are able to perform
secure QKD in order for $S$ to send secret classical information to those
consumers. \ Because any protocol for QKD including BB84\cite{bb84} is
effective, we do not specify QKD protocols. Also assume that all $C_{m}$ and 
$S$ share a set of many spin chains in the ground state $|g\rangle $. Now
let me explain the protocol explicitly. The protocol is composed of the
following six steps.

\bigskip

(1) $S$ performs a local projective measurement of observable $U_{S\text{ }}$%
for the ground state $|g\rangle $. Assume that $S$ obtains the measurement
result $\mu $. $S$ must input energy $E_{S}$ on average to the spin chain in
order to achieve that local measurement. $E_{S}$ is evaluated as 
\begin{equation*}
E_{S}=\sum_{\mu =0,1}\langle g|P_{S}\left( \mu \right) HP_{S}\left( \mu
\right) |g\rangle .
\end{equation*}

\bigskip

(2) $S$ authenticates $C_{m}$ by use of common secret short keys $k$.

\bigskip

(3) $S$ and authenticated $C_{m}$'s $~$generate and share sufficiently-long
pseudo-random secret keys $K$ via a protocol for QKD.

\bigskip

(4) $\ S$ encodes the measurement results $\mu $ by use of $K$ and sends to
authenticated $C_{m}$'s.

\bigskip

(5) $C_{m}$ decodes the measurement reults $\mu $ by use of $K$.

\bigskip

(6) $C_{m}$'s perform a local unitary operation $V_{m}\left( \mu \right) $
in Eq.(\ref{vb}) to their qubits, depending on the value of $\mu $.~Each $%
C_{m}$ obtains energy output $E_{m}~$on average from the spin chains in this
process. $E_{m}$ is given by

\begin{equation}
E_{m}=\frac{1}{2}\left[ \sqrt{\xi _{m}^{2}+\eta _{m}^{2}}-\xi _{m}\right] ,
\label{em}
\end{equation}%
where%
\begin{equation*}
\xi _{m}=\langle g|U_{m}^{\dag }HU_{m}|g\rangle ,
\end{equation*}%
\begin{equation*}
\eta _{m}=\langle g|U_{S}\dot{U}_{m}|g\rangle .
\end{equation*}

\bigskip

After step (6), the quantum state is given by%
\begin{equation}
\rho _{QED}^{(6)}=\sum_{\mu =0,1}\left( \prod_{m}V_{m}\left( \mu \right)
\right) P_{S}\left( \mu \right) |g\rangle \langle g|P_{S}\left( \mu \right)
\left( \prod_{m}V_{m}^{\dag }\left( \mu \right) \right) .  \label{5}
\end{equation}%
This QED protocol is robust against impersonation attack. Let us imagine
that an illegal consumer Derick appears at site $n_{D}$, who does not have $%
k $ and attempts to get energy from spin chains. Then we can conclude that
Derick does not obtain but give energy to the spin chains. The reason is
following. Because Derick cannot get no information about $\mu $, Derick
makes randomly two local operations $V_{D}\left( 0\right) $ and $V_{D}\left(
1\right) $ given by 
\begin{equation*}
V_{D}\left( \mu \right) =I\cos \theta +i\left( -1\right) ^{\mu }\vec{n}%
_{D}\cdot \vec{\sigma}_{n_{D}}\sin \theta .
\end{equation*}%
Then, instead of Eq.(\ref{5}), the final state becomes%
\begin{equation*}
\rho _{D}=\frac{1}{2}\sum_{\mu ,\mu ^{\prime }}V_{D}\left( \mu ^{\prime
}\right) \left( \prod_{m}V_{m}^{\dag }\left( \mu \right) \right) P_{S}\left(
\mu \right) |g\rangle \langle g|P_{S}\left( \mu \right) \left(
\prod_{m}V_{m}^{\dag }\left( \mu \right) \right) V_{D}^{\dag }\left( \mu
^{\prime }\right) .
\end{equation*}%
Evaluation of the average localized energy around Derick is straightforward
and gives a positive value such that 
\begin{equation*}
\limfunc{Tr}\left[ \rho _{D}H_{D}\right] =\frac{1}{2}\sum_{\mu ^{\prime
}=0,1}\langle g|V_{D}^{\dag }\left( \mu ^{\prime }\right) HV_{D}\left( \mu
^{\prime }\right) |g\rangle >0.
\end{equation*}%
Here we have used

\begin{equation*}
\left[ ~\prod_{m}V_{m}\left( \mu \right) ,~V_{D}^{\dag }\left( \mu ^{\prime
}\right) H_{D}V_{D}\left( \mu ^{\prime }\right) \right] =0,
\end{equation*}%
\begin{equation*}
\left[ P_{S}\left( \mu \right) ,~V_{D}^{\dag }\left( \mu ^{\prime }\right)
H_{D}V_{D}\left( \mu ^{\prime }\right) \right] =0,
\end{equation*}%
and%
\begin{equation*}
\langle g|V_{D}^{\dag }\left( \mu ^{\prime }\right) H_{D}V_{D}\left( \mu
^{\prime }\right) |g\rangle =\langle g|V_{D}^{\dag }\left( \mu ^{\prime
}\right) HV_{D}\left( \mu ^{\prime }\right) |g\rangle .
\end{equation*}%
Because the value of $\limfunc{Tr}\left[ \rho _{D}H_{D}\right] $ is
positive, Derick must input energy on average to the spin chains without
gain.

Finally we add a comment that it is possible to array an infinite number of
consumers in the most dense distribution by putting consumers at $n=5m$ for
\thinspace nonzero integer $m$. The total amount of energy gain by the
consumers is defined by%
\begin{equation}
E_{C}=-\sum_{m\neq 0}\limfunc{Tr}\left[ \rho _{QED}^{(6)}H_{C_{m}}\right] =%
\frac{1}{2}\sum_{m\neq 0}\left[ \sqrt{\xi _{m}^{2}+\eta _{m}^{2}}-\xi _{m}%
\right] .  \label{c}
\end{equation}

\bigskip

\section{Local Cooling by Energy Supplier}

\bigskip

\bigskip

\bigskip

In this section, we discuss a relation between QED and a local cooling
process by the energy supplier $S$ of QED. In step (1) of the previous QED
protocol, $S$ must deposit energy $E_{A}$ to the spin chain. Let us imagine
that $S$ stops the protocol soon after step (1) and attempts to completely
withdraw $E_{A}~$by local operations. By a similar argument in \cite{hotta1}%
, \ it is shown that this attempt never succeeds. In step (1), $S$ breaks
entanglement between $S$'s qubit and other qubits and the entanglement
cannot be recovered only by local operations. (Of course, for a long time
interval beyond the short time scale that we have considered, local cooling
is naturally expected to make residual energy approaching zero by an assist
of dynamical evolution induced by nonlocal Hamiltonians. The time evolution
is able to recover the entanglement broken by $S$.) Hence, there exists
nonvanishing residual energy $E_{r}$ of the local cooling. Though explicit
values of $E_{r}$ can be obtained for a special class of spin chain systems,
including the Ising spin chain analyzed in the next section, the evaluation
of $E_{r}$ is not so easy for general spin chains. However, $E_{C}$ in Eq.(%
\ref{c}) generally gives a lower bound of $E_{r}$.

The reason is following. Let us consider a general local operation of $S$,
which is expressed by use of $\mu $-dependent measurement operators $%
M_{S}(\alpha ,\mu )$ satisfying%
\begin{equation*}
\sum_{\alpha }M_{S}^{\dag }(\alpha ,\mu )M_{S}(\alpha ,\mu )=I.
\end{equation*}%
Then the quantum state after that local cooling by $S$ is given by%
\begin{equation}
\rho _{c}=\sum_{\mu ,\alpha }M_{S}(\alpha ,\mu )P_{S}\left( \mu \right)
|g\rangle \langle g|P_{S}\left( \mu \right) M_{S}^{\dag }(\alpha ,\mu ).
\label{31}
\end{equation}%
The residual energy $E_{r}$ is evaluated as

\begin{equation}
E_{r}=\min_{\left\{ M_{S}(\alpha ,\mu )\right\} }\limfunc{Tr}\left[ \rho
_{c}H_{S}\right] ,  \label{40}
\end{equation}%
where $H_{S}$ is the energy density of $S$ given by 
\begin{equation*}
H_{S}=\sum_{n=-1}^{1}T_{n}.
\end{equation*}%
The key point is that the value of $E_{r}$ can be calculated from the
quantum state of QED. If $S$ performs the above local cooling after the end
of the QED protocol, the quantum state is transformed from that in Eq.(\ref%
{5}) to

\begin{equation*}
\rho _{QED}^{(C)}=\sum_{\mu \alpha }M_{S}(\alpha ,\mu )\left( \prod_{m\neq
0}V_{m}\left( \mu \right) \right) P_{S}\left( \mu \right) |g\rangle \langle
g|P_{S}\left( \mu \right) \left( \prod_{m\neq 0}V_{m}^{\dag }\left( \mu
\right) \right) M_{S}^{\dag }(\alpha ,\mu ).
\end{equation*}%
Here it is easily proven that 
\begin{equation*}
\limfunc{Tr}\left[ \rho _{QED}^{(C)}H_{S}\right] =\limfunc{Tr}\left[ \rho
_{c}H_{S}\right]
\end{equation*}%
because $M_{S}(\alpha ,\mu )$ and $V_{m}\left( \mu \right) $ commute with
each other. Thus $E_{r}$ is rewritten as

\begin{equation}
E_{r}=\min_{\left\{ M_{S}(\alpha ,\mu )\right\} }\limfunc{Tr}\left[ \rho
_{QED}^{(C)}H_{S}\right] .  \label{9}
\end{equation}%
It is stressed that the following relation should hold because of
nonnegativity of $H.$ 
\begin{equation}
\limfunc{Tr}\left[ \rho _{QED}^{(C)}H\right] =\limfunc{Tr}\left[ \rho
_{QED}^{(C)}H_{S}\right] +\sum_{m\neq 0}\limfunc{Tr}\left[ \rho
_{QED}^{(C)}H_{C_{m}}\right] \geq 0.  \label{7}
\end{equation}%
Moreover, it is shown that%
\begin{equation}
\sum_{m\neq 0}\limfunc{Tr}\left[ \rho _{QED}^{(C)}H_{C_{m}}\right]
=\sum_{m\neq 0}\limfunc{Tr}\left[ \rho _{QED}^{(6)}H_{C_{m}}\right] =-E_{C},
\label{8}
\end{equation}%
where $\rho _{QED}^{(6)}$ is the quantum state in Eq.(\ref{5}). From Eq.(\ref%
{7}) and Eq.(\ref{8}), we obtain

\begin{equation*}
\limfunc{Tr}\left[ \rho _{QED}^{(C)}H_{S}\right] -E_{C}\geq 0.
\end{equation*}%
This gives a relation what we want by taking account of Eq.(\ref{9}):

\begin{equation}
E_{r}\geq E_{C}.  \label{10}
\end{equation}%
Thus it has been proven that $E_{C}$ in Eq.(\ref{c}) gives a lower bound of $%
E_{r}$.

There may be a question whether the bound in Eq.(\ref{10}) is achievable or
not. However, this is very nontrivial. One of the neccesary conditions is to
achieve the equality of Eq.(\ref{7}) even if negative energy density appears
in some region. Though the answer is not known for spin chain systems,
equality of a similar relation does not hold for a free field in two
dimensional spacetime \cite{hotta2}. Hence, it might be impossible to attain
the bound in Eq.(\ref{10}).

\section{Ising Chain Analysis}

\bigskip

\bigskip

In this section, we demonstrate QET and QED protocols in Ising models
without spontaneous symmetry breaking. Detailed properties of the model can
be seen in \cite{Ising}-\cite{P}. Let us write the Hamiltonian as

\begin{equation*}
H=-h\sum_{n=-\infty }^{\infty }\sigma _{n}^{z}-J\sum_{n=-\infty }^{\infty
}\sigma _{n}^{x}\sigma _{n+1}^{x}-E_{g},
\end{equation*}%
where $h\geq J>0$ and $E_{g}$ is a constant which shifts the eigenvalue of
the ground state $|g\rangle $ to zero:%
\begin{equation*}
H|g\rangle =0,
\end{equation*}%
and $\sigma _{n}^{z}$ and $\sigma _{n}^{x}$ are Pauli matrices at site $n$.

Let us introduce a parameter $\lambda =J/h(\leq 1)$. When $\lambda =1$, the
system becomes the critical Ising model. The energy density operator at site 
$n$ is defined by

\begin{equation*}
T_{n}=-h\sigma _{n}^{z}-\frac{J}{2}\sigma _{n}^{x}\left( \sigma
_{n+1}^{x}+\sigma _{n-1}^{x}\right) -\epsilon ,
\end{equation*}%
where $\epsilon $ is a real constant to satisfy $\langle g|T_{n}|g\rangle =0$%
. The Hamiltonian can be expressed as a sum of $T_{n}$:

\begin{equation*}
H=\sum_{n}T_{n}.
\end{equation*}%
The system can be mapped into a Fermionic system and solved analytically\cite%
{Ising}. It is shown that correlation functions are evaluated, for exmaple,
as 
\begin{equation*}
\langle g|\sigma _{n}^{x}|g\rangle =\langle g|\sigma _{n}^{y}|g\rangle =0,
\end{equation*}%
\begin{equation*}
\langle g|\sigma _{n}^{z}|g\rangle =G(0),
\end{equation*}%
\begin{equation*}
\langle g|\sigma _{0}^{x}\left( \sigma _{1}^{x}+\sigma _{-1}^{x}\right)
|g\rangle =2G(-1),
\end{equation*}%
\begin{eqnarray*}
\langle g|\sigma _{m}^{y}\sigma _{m+n}^{y}|g\rangle  &=&\Delta (n) \\
&=&\left\vert 
\begin{array}{cccc}
G(1) & G(0) & \cdots  & G(2-n) \\ 
G(2) & G(1) & \cdots  & G(3-n) \\ 
\vdots  & \vdots  & \ddots  & \vdots  \\ 
G(n) & G(n-1) & \cdots  & G(1)%
\end{array}%
\right\vert ,
\end{eqnarray*}%
where the function $G(n)$ is defined by%
\begin{eqnarray*}
G(n) &=&L(n)+\lambda L(n+1), \\
L(n) &=&\frac{1}{\pi }\int_{0}^{\pi }\frac{\cos \left( kn\right) }{\sqrt{%
1+\lambda ^{2}+2\lambda \cos k}}dk.
\end{eqnarray*}%
A relation%
\begin{equation}
\langle g|\sigma _{0}^{y}\left( \sigma _{1}^{x}+\sigma _{-1}^{x}\right)
|g\rangle =0  \label{33}
\end{equation}%
is also obtained explicitly. For the case with $\lambda =1$, $G(n)$ is
calculated as

\begin{equation*}
G(n)=\frac{2}{\pi }\frac{\left( -1\right) ^{n}}{2n+1}.
\end{equation*}%
Then the correlation functions are given by

\begin{equation*}
\langle g|\sigma _{n}^{z}|g\rangle =\frac{2}{\pi },
\end{equation*}%
\begin{equation}
\langle g|\sigma _{0}^{x}\left( \sigma _{1}^{x}+\sigma _{-1}^{x}\right)
|g\rangle =\frac{4}{\pi },  \label{34}
\end{equation}

\begin{equation*}
\langle g|\sigma _{m}^{y}\sigma _{m+n}^{y}|g\rangle =-\left( \frac{2}{\pi }%
\right) ^{n}\frac{2^{2n(n-1)}h(n)^{4}}{\left( 4n^{2}-1\right) h(2n)},
\end{equation*}%
where

\begin{equation*}
h(n)=\prod_{k=1}^{n-1}k^{n-k}.
\end{equation*}%
The asymptotic behavior of $\Delta (n)$ for large $n$ is given by 
\begin{equation}
\Delta (n\sim \infty )\sim -\frac{1}{4}e^{1/4}2^{1/12}c^{-3}n^{-9/4},
\label{20}
\end{equation}%
where the constant $c$ is evaluated as $c\sim 1.28$ \cite{P}. \ 

Now let us first consider QET. To specify the protocol, we set 
\begin{equation}
U_{A}=\sigma _{n_{A}}^{y},  \label{50}
\end{equation}%
and%
\begin{equation}
U_{B}=\sigma _{n_{B}}^{x}.  \label{51}
\end{equation}%
The energy input $E_{A}$ by the energy supplier \thinspace of Eq.(\ref{12})
is evaluated as

\begin{equation*}
E_{A}=\frac{1}{2}\langle g|\sigma _{n_{A}}^{y}H\sigma _{n_{A}}^{y}|g\rangle
=h\langle g|\sigma _{n_{A}}^{z}|g\rangle +J\langle g|\sigma
_{n_{A}}^{x}\left( \sigma _{n_{A}+1}^{x}+\sigma _{n_{A}-1}^{x}\right)
|g\rangle =hG(0)+2JG(-1),
\end{equation*}%
where we have used a relation as

\begin{equation*}
P_{S}\left( \mu \right) =\frac{1}{2}\left[ I+\left( -1\right) ^{\mu }\sigma
_{n_{A}}^{y}\right] .
\end{equation*}%
For the case with $\lambda =1$, $E_{A}$ is given by 
\begin{equation*}
E_{A}=\frac{6}{\pi }h.
\end{equation*}%
The coefficient $\xi $ in Eq.(\ref{15}) is evaluated as

\begin{equation*}
\xi =\langle g|\sigma _{n_{B}}^{x}H\sigma _{n_{B}}^{x}|g\rangle =2h\langle
g|\sigma _{n_{B}}^{z}|g\rangle =2hG(0).
\end{equation*}%
The time-derivative operator of $U_{B}$ is given by

\begin{equation*}
\dot{U}_{B}=i\left[ -h\sigma _{n_{B}}^{z},~\sigma _{n_{B}}^{x}\right]
=2h\sigma _{n_{B}}^{y}.
\end{equation*}%
Hence, the value of $\eta $ in Eq.(\ref{16}) ~is calculated as

\begin{equation*}
\eta =\langle g|U_{A}\dot{U}_{B}|g\rangle =2h\langle g|\sigma
_{n_{A}}^{y}\sigma _{n_{B}}^{y}|g\rangle =2h\Delta (\left\vert
n_{A}-n_{B}\right\vert ).
\end{equation*}%
From these values of $\xi $ and $\eta $, the energy output $E_{B}$ of QET is
computed as

\begin{equation*}
E_{B}=\frac{2h}{\pi }\left[ \sqrt{1+\left( \frac{\pi }{2}\Delta (\left\vert
n_{A}-n_{B}\right\vert )\right) ^{2}}-1\right] .
\end{equation*}%
It should be stressed that nonvanishing values of $E_{B}$ \ are obtained for
general values of $\lambda $, including noncritical models. It is noted that
the evaluation of $E_{B}$ is simple for the critical Ising case with $%
\lambda =1$. The asymptotic value of $E_{B}$ with $\lambda =1$ is obtained
for $\left\vert n_{B}-n_{A}\right\vert \sim \infty $ from Eq.(\ref{20}) as
follows.%
\begin{align}
E_{B}& \sim \frac{h}{\pi }\left( \frac{\pi }{2}\Delta (\left\vert
n_{B}-n_{A}\right\vert )\right) ^{2}  \notag \\
& \sim h\frac{\pi }{64}\sqrt{e}2^{1/6}c^{-6}\left\vert
n_{B}-n_{A}\right\vert ^{-9/2}.  \label{21}
\end{align}

Next let us consider QED with an infinite number of consumers in the most
dense distribution. When $\lambda =1$, the total amount of energy transfer $%
E_{C}$ in Eq.(\ref{c}) can be evaluated explicitly as%
\begin{eqnarray}
E_{C} &=&\frac{2h}{\pi }\sum_{m\neq 0}\left[ \sqrt{1+\left( \frac{\pi }{2}%
\Delta (5|m|)\right) ^{2}}-1\right]  \notag \\
&\sim &6.2\times 10^{-5}h.  \label{ecc}
\end{eqnarray}%
As discussed in section 4, this value gives a lower bound of $E_{r}$ of
local cooling by $S$. In this solvable model, we can check explicitly the
relation in Eq.(\ref{10}). The minimization of $E_{r}$ in Eq.(\ref{40})
among local operations is possible.

For general values of $\lambda $, the localized \ energy $H_{S}$ is
expicitly written as

\begin{align*}
H_{S}& =-h\sigma _{0}^{z}-J\sigma _{0}^{x}\left( \sigma _{1}^{x}+\sigma
_{-1}^{x}\right) \\
& -\frac{J}{2}\sigma _{1}^{x}\sigma _{2}^{x}-\frac{J}{2}\sigma
_{-1}^{x}\sigma _{-2}^{x}-J\left( \sigma _{-1}^{z}+\sigma _{1}^{z}\right)
-3\epsilon .
\end{align*}%
It is noted that the following relation holds for $\rho _{c}$ in Eq. (\ref%
{31}). 
\begin{eqnarray*}
&&\limfunc{Tr}\left[ \rho _{c}\left( -\frac{J}{2}\sigma _{1}^{x}\sigma
_{2}^{x}-\frac{J}{2}\sigma _{-1}^{x}\sigma _{-2}^{x}-J\left( \sigma
_{-1}^{z}+\sigma _{1}^{z}\right) -3\epsilon \right) \right] \\
&=&\langle g|\left( -\frac{J}{2}\sigma _{1}^{x}\sigma _{2}^{x}-\frac{J}{2}%
\sigma _{-1}^{x}\sigma _{-2}^{x}-J\left( \sigma _{-1}^{z}+\sigma
_{1}^{z}\right) -3\epsilon \right) |g\rangle .
\end{eqnarray*}%
By use of the above relation, we are able to manipulate as follows.%
\begin{eqnarray*}
\limfunc{Tr}\left[ \rho _{c}H_{S}\right] &=&\limfunc{Tr}\left[ \rho
_{c}\left( -h\sigma _{0}^{z}-J\sigma _{0}^{x}\left( \sigma _{1}^{x}+\sigma
_{-1}^{x}\right) \right) \right] \\
&&+\langle g|\left( -\frac{J}{2}\sigma _{1}^{x}\sigma _{2}^{x}-\frac{J}{2}%
\sigma _{-1}^{x}\sigma _{-2}^{x}-J\left( \sigma _{-1}^{z}+\sigma
_{1}^{z}\right) -3\epsilon \right) |g\rangle \\
&=&\langle g|\left( h\sigma _{0}^{z}+J\sigma _{0}^{x}\left( \sigma
_{1}^{x}+\sigma _{-1}^{x}\right) \right) |g\rangle -\limfunc{Tr}\left[ \rho
_{c}\left( h\sigma _{0}^{z}+J\sigma _{0}^{x}\left( \sigma _{1}^{x}+\sigma
_{-1}^{x}\right) \right) \right] +\langle g|H_{S}|g\rangle .
\end{eqnarray*}%
By substituting $\langle g|H_{S}|g\rangle =0$ and\ 
\begin{equation*}
\langle g|\left( h\sigma _{0}^{z}+J\sigma _{0}^{x}\left( \sigma
_{1}^{x}+\sigma _{-1}^{x}\right) \right) |g\rangle =hG(0)+2JG(-1),
\end{equation*}%
it is obtained that

\begin{equation*}
\limfunc{Tr}\left[ \rho _{c}H_{S}\right] =hG(0)+2JG(-1)-\limfunc{Tr}\left[
\rho _{c}\left( h\sigma _{0}^{z}+J\sigma _{0}^{x}\left( \sigma
_{1}^{x}+\sigma _{-1}^{x}\right) \right) \right] .
\end{equation*}%
Here it is useful to write $P_{S}\left( \mu \right) $ as follows.

\begin{equation}
P_{S}\left( \mu \right) =|\mu ,n=0\rangle \langle \mu ,n=0|\prod_{n\neq
0}I_{n}=\frac{1}{2}\left[ I+\left( -1\right) ^{\mu }\sigma _{0}^{y}\right] .
\label{peq}
\end{equation}%
After messy calculations using Eq. (\ref{peq}), we obtain

\begin{equation}
\limfunc{Tr}\left[ \rho _{c}H_{S}\right] =hG(0)+2JG(-1)-h\limfunc{Tr}\left[
\rho _{S}\sigma _{0}^{z}\right] .  \label{feq}
\end{equation}%
In general, the following inequality holds.%
\begin{equation*}
1\geq \limfunc{Tr}\left[ \rho _{S}\sigma _{0}^{z}\right] .
\end{equation*}%
The equality is attained by measurement operators as

\begin{eqnarray*}
M_{A}(\mu &=&0)=\frac{1}{2}\left[ 
\begin{array}{cc}
1 & -i \\ 
1 & i%
\end{array}%
\right] , \\
M_{A}(\mu &=&1)=\frac{1}{2}\left[ 
\begin{array}{cc}
1 & i \\ 
1 & -i%
\end{array}%
\right] ,
\end{eqnarray*}%
without the $\alpha $ degree of freedom. These lead to the final result as
followed.%
\begin{equation*}
E_{r}=h\left( G(0)-1\right) +2JG(-1).
\end{equation*}%
When $\lambda =1$, the value of $E_{r}$ is given by

\begin{equation*}
E_{r}=\left( \frac{6}{\pi }-1\right) J\sim 0.91J.
\end{equation*}%
Because evaluation of the total amount of energy transfer $E_{C}$ in Eq.(\ref%
{c}) is difficult, comparison between $E_{C}$ and $E_{r}$ is not performed
easily. However, by taking $\lambda =1$, the comparison is possible as
follows. From Eq. (\ref{ecc}), Eq.(\ref{10}) is verified because $%
0.91>6.2\times 10^{-5}$. In this case, the bound of Eq.(\ref{10}) is not so
tight. This result depends on the choice of $U_{A}$ and $U_{B}$ in Eq.(\ref%
{50}) and Eq.(\ref{51}). If we change the choice, the bound may become
tighter.

\bigskip

\section{Conclusion and Discussions}

\bigskip

In this paper, a protocol for QED is proposed, in which many consumers can
simultaneously extract energy from spin chains by use of common secret keys
shared by an energy supplier. In this protocol, what consumers need for
energy gain is just classical information about measurement results. Hence,
dissipation process in energy transportation via channels can be neglected.
This protocol for QED is robust against impersonation. An adversary, who
does not have common secret keys and attempts to get energy, cannot obtain
but give energy to the spin chains. We have also pointed out that the total
amount of energy distributed via QED gives a lower bound of residual energy
in a supplier's local cooling for a state excited by the supplier's
measurement. \ Finally, QET and QED protocols have been studied in the Ising
spin chain model. Amount of energy transmission is explicitly evaluated
depending on distance from the supplier. Finally, the lower bound of Eq.(\ref%
{10}) has been explicitly checked.

For practical situations for QED, there remain some open problems of QED.
They are listed below.

One of them is related with energy dissipation. It is expected that
dissipation effects in the energy transport of QED are severely suppressed
even if a zero-temperature uncontrolled environment is coupled with the spin
chain. What those consumers need for energy gain is just classical
information about the measurement result without receiving energy directly
from the supplier. This aspects is quite a contrast to ordinary energy
transmission in the spin chains. In the transportation, excitations as
energy carriers in the spin chain get gradually annihilated dependent on
environment interaction properties. Detailed analysis of comparison between
QED and ordinary energy transports in spin chains is an interesting problem.
However, the analysis is out of scope of this paper and will be discuss
elsewhere. The finite temperature effect is also considered to be of
importance for QED.

In the QED protocol, we fix a unitary operation of $B$ which takes a form in
Eq. (\ref{vb}). However, the optimal local operation of $B$ to extract
maximum energy from the spin chain is not obtained yet. The optimal
operation is crucial for analyses of local cooling because the extracted
energy gives a lower bound of residual energy of local cooling by the
supplier. \ \ 

Realistic implemetation proposals of QED are not reported yet. However,
there are possible candidates. One of them might be the carbon nanotube.
Carbon nanotubes are able to contain fullerenes with atoms inside. The
fullerenes acquire spins by doping some atoms. By laying the fullerenes
side-by-side in a nanotube, a spin chain might be constructed and useful to
check QED. Quantum dots and SQUID qubit systems might allow to create spin
chains for QED.

\bigskip

\bigskip

\textbf{Acknowledgments}\newline

\bigskip

This research was partially supported by the SCOPE project of the MIC.

\bigskip

\end{document}